\documentclass[twocolumn,aps,floatfix,showpacs,superscriptaddress]{revtex4}
\usepackage{amsmath,amssymb,eucal,graphicx,bm}
\begin{document}
\title{Scaling Exponents for Ordered Maxima}
\author{E.~Ben-Naim}
%\email{ebn@lanl.gov}
\affiliation{Theoretical Division and Center for Nonlinear Studies,
Los Alamos National Laboratory, Los Alamos, New Mexico 87545, USA}
\author{P.~L.~Krapivsky}
%\email{paulk@bu.edu}
\affiliation{Department of Physics, Boston University, Boston,
Massachusetts 02215, USA}
\affiliation{Institut de Physique Th\'eorique, Universit\'e Paris-Saclay, 
CEA and CNRS,
91191 Gif-sur-Yvette, France}
\author{N.~W.~Lemons}
%\email{nlemons@lanl.gov}
\affiliation{Theoretical Division and Center for Nonlinear Studies,
Los Alamos National Laboratory, Los Alamos, New Mexico 87545, USA}
\begin{abstract}
  We study extreme value statistics of multiple sequences of random
  variables.  For each sequence with $N$ variables, independently
  drawn from the same distribution, the running maximum is defined as
  the largest variable to date. We compare the running maxima of $m$
  independent sequences, and investigate the probability $S_N$ that
  the maxima are perfectly ordered, that is, the running maximum of
  the first sequence is always larger than that of the second
  sequence, which is always larger than the running maximum of the
  third sequence, and so on.  The probability $S_N$ is universal: it
  does not depend on the distribution from which the random variables
  are drawn.  For two sequences, $S_N\sim N^{-1/2}$, and in general,
  the decay is algebraic, $S_N\sim N^{-\sigma_m}$, for large $N$. We
  analytically obtain the exponent $\sigma_3\cong 1.302931$ as root of
  a transcendental equation. Furthermore, the exponents $\sigma_m$
  grow with $m$, and we show that $\sigma_m \sim m$ for large $m$.
\end{abstract}
\maketitle

\section{Introduction}

The theory of extreme values is a well-developed area of statistics
and probability theory \cite{llr,eig,rse,sir,abn,vbn}. Extreme values
such as the maximal and the minimal data points are important features
of a dataset. Statistics of extreme events play a key role in a host
of data rich subjects including climate science \cite{rp,nmt,whk},
geophysics \cite{vzm,win,sdt,wzxw}, and economics
\cite{ekm,bp,syn,wbk}.

The running maximum, defined as the largest variable to date in a
sequence of variables, is a central quantity in extreme-value
statistics. This quantity evolves rather slowly: the number of times
it changes typically grows logarithmically with the number of random
variables \cite{abn,vbn}. Consequently, persistence \cite{msb,dhz,bms} or
first-passage properties \cite{sr} involving the running maxima often
exhibit power-law dependence on the number of variables
\cite{bk13,mb,bk14}. Methods and concepts from statistical physics
provide a powerful tool for obtaining the nontrivial scaling exponents
that characterize such power-law behaviors \cite{krb}.

This investigation is motivated by a recent letter \cite{bk}
concerning maximal positions of random walks. It was reported that the
probability that the maxima of multiple random walks remain perfectly
ordered decays as a power law with the number of steps \cite{bk,jrf},
and that the corresponding decay exponents are generally nontrivial.

\begin{figure}[t]
\vspace{.2in}
\includegraphics[width=0.4\textwidth]{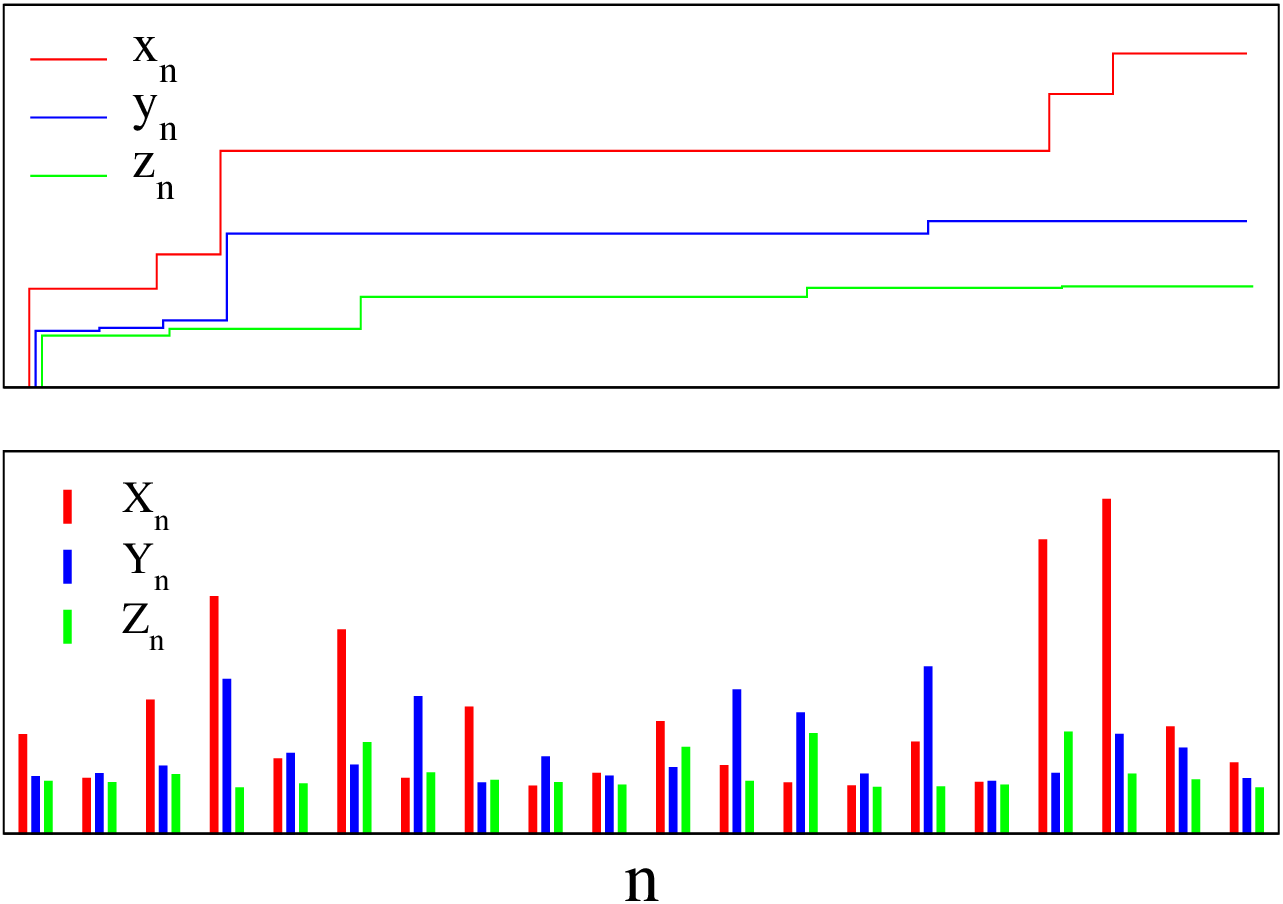}
\caption{Illustration of perfectly ordered maxima. The lower panel
  shows three sequences of random variables (bars), and the upper
  panel displays the corresponding running maxima (staircases).  The
  latter sequences are perfectly ordered.  The leading sequence is
  shifted slightly to the left and the trailing sequence is shifted
  slightly to the right.}
\label{fig-order}
\end{figure}

Here we study the same probability for {\em uncorrelated} random
variables.  Figure \ref{fig-order} shows three sequences of random
variables (bottom panel) and the corresponding running maxima (top
panel). The running maxima form three staircases. We are interested in
the probability that these running maxima remain perfectly ordered, or
equivalently, that the staircases do not intersect even once. When the
random variables are uncorrelated, this probability is universal: it
is completely independent of the distribution from which the variables
are drawn.

The probability $S_N$ that the three staircases are ordered decays
algebraically with the number of random variables.  Interestingly, the
decay exponent is nontrivial:
\begin{equation}
S_N\sim N^{-\sigma},\quad\text{with}\quad \sigma=1.302931.
\end{equation}
We obtain this exponent analytically. In the general case, we obtain
upper and lower bounds implying that the exponent grows linearly with
the number of independent sequences. Interestingly, this family of
exponents gives a good approximation for the first-passage exponents
found for maxima of random walks.

The rest of this paper is organized as follows. In section II, we
investigate statistics of perfectly ordered maxima. We start with two
sequences for which the analysis is straightforward.  We then present
theoretical results for the nontrivial case of three sequences. For
the general case, we obtain the leading asymptotic behavior of the
scaling exponents.  In section III, we treat a related question
concerning partially ordered maxima, and show that the exponents
obtained in section II are part of an infinite set of families of
scaling exponents. We conclude with a summary and a discussion of
related open problems (Section IV). The appendix details technical
derivations used in the three-sequence case.

\section{Perfect Order}

We study extreme values of multiple sequences of $N$ uncorrelated
random variables. Let us first consider one sequence of random
variables 
\begin{equation}
\label{sequence}
\{X_1,X_2,\ldots,X_N\},
\end{equation}
where each of the variables $X_i$ is drawn from the same
distribution. We restrict our attention to continuous distributions
for which there are no ties: $X_i\neq X_j$ for all $i\neq j$.  The
running maximum $x_n$ is defined as the largest variable to date
\begin{equation}
\label{record-def}
x_n=\text{max}\{X_1,X_2,\ldots,X_n\}\, ,
\end{equation}
with $n=1,2,\ldots,N$. Overall there are $N$ maxima.  These maxima are
monotonically increasing $x_{n+1}\geq x_n$ for all $n$, and form the
sequence $\{x_1,x_2,\ldots,x_N\}$. Figure \ref{fig-order} illustrates
that maxima are {\em correlated} stochastic variables: by the
definition \eqref{record-def}, a running maxima involves memory of all
preceding random variables.

In this study, we consider $m$ independent sequences of random
variables such as \eqref{sequence} and their corresponding maxima
defined by \eqref{record-def} and ask: what is the probability that
the maxima remain perfectly ordered?

\subsection{Two Sequences}

We start with two sequences. The second sequence of random variables,
$\{Y_1,Y_2,\ldots,Y_N\}$, is drawn from the same distribution as the
first sequence \eqref{sequence}. The running maxima for the second
sequence are again defined by $y_n=\text{max}\{Y_1,Y_2,\ldots,Y_n\}$
for $n=1,2,\ldots,N$. We are interested in the probability $S_N$ that
the first set of maxima is always larger then the second set:
\begin{equation}
\label{condition2}
x_n > y_n, \qquad  n=1,2,\ldots,N. 
\end{equation}
We term $S_N$ the ``survival'' probability since the condition
\eqref{condition2} defines a first-passage process \cite{sr}.

The survival probability $S_N$ obeys the {\em closed} recursion relation 
\begin{equation}
\label{SN-2-rec}
S_{N}=S_{N-1}\left(1-\frac{1}{2N}\right),
\end{equation}
subject to the ``initial'' condition $S_0=1$. To appreciate
\eqref{SN-2-rec} let us combine the two sets of $N$ random variables
into a larger set of $2N$ variables. Since all of these random
variables are drawn from the same distribution, each variable is
equally likely to be the largest. In particular, the variable $Y_N$ is
the largest with probability $\frac{1}{2N}$. The probability that the
leading sequence remains in the lead at the $N$th step is therefore
$1-\frac{1}{2N}$.

Using Eq.~\eqref{SN-2-rec} one gets $S_1=\frac{1}{2}$,
\hbox{$S_2=\frac{3}{8}$}, $S_3=\frac{5}{16}$, and generally
\begin{equation}
\label{SN-2-sol}
S_N=\binom{2N}{N}\,2^{-2N}.
\end{equation}
Importantly, this probability is {\em universal} as it holds
regardless of the distribution from which the random variables are
drawn. There are two requirements for equation \eqref{SN-2-sol} to
hold: (i) all $2N$ variables must be independent and identically
distributed, and (ii) the probability distribution that governs these
random variables is continuous so that there are no ties.

To obtain the asymptotic behavior for large $N$, we use the Stirling
 formula \hbox{$N!\simeq (2\pi N)^{-1/2}(N/e)^N$} and equation
 \eqref{SN-2-sol}.  We thus find that the survival probability decays
 algebraically,
\begin{equation}
\label{SN-2-decay}
S_N\simeq \pi^{-1/2}\, N^{-1/2}
\end{equation}
for large $N$.  As shown below, the survival probability has a similar
algebraic decay in the general case, except that the decay exponent
depends on the number of sequences.

The same probability distribution \eqref{SN-2-sol} arises in the
context of discrete time random walks. Consider two walks that start
at the origin. The probability $S_N$ that the positions remain
ordered, $x_1(n)>x_2(n)$ for $1\leq n\leq N$, is given by
Eq.~\eqref{SN-2-sol}. This result, known as the Sparre Andersen
theorem \cite{SA54,F68,SM}, remains valid {\em regardless} of the step
distribution, e.g., it holds for Levy walks with diverging average
step length \cite{pl}.

We also remark that the random variables $X_n$ are uncorrelated, e.g.
$\langle X_iX_j\rangle=\langle X_i \rangle \langle X_j\rangle$ for all
$i\neq j$; further, there are no correlations between the sequences
$X_n$ and $Y_n$.  As a result, the probability $\Pi_N$ that the actual
random variables are always ordered, $X_n>Y_n$ for all $n$, is purely
exponential $\Pi_N=2^{-N}$. In view of the much slower algebraic decay
\eqref{SN-2-decay}, we conclude that ordered sequences of random
variables are much less likely than ordered sequences of maxima. 

\subsection{Three Sequences}
\label{sec:3}

\begin{figure}[t]
\includegraphics[width=0.425\textwidth]{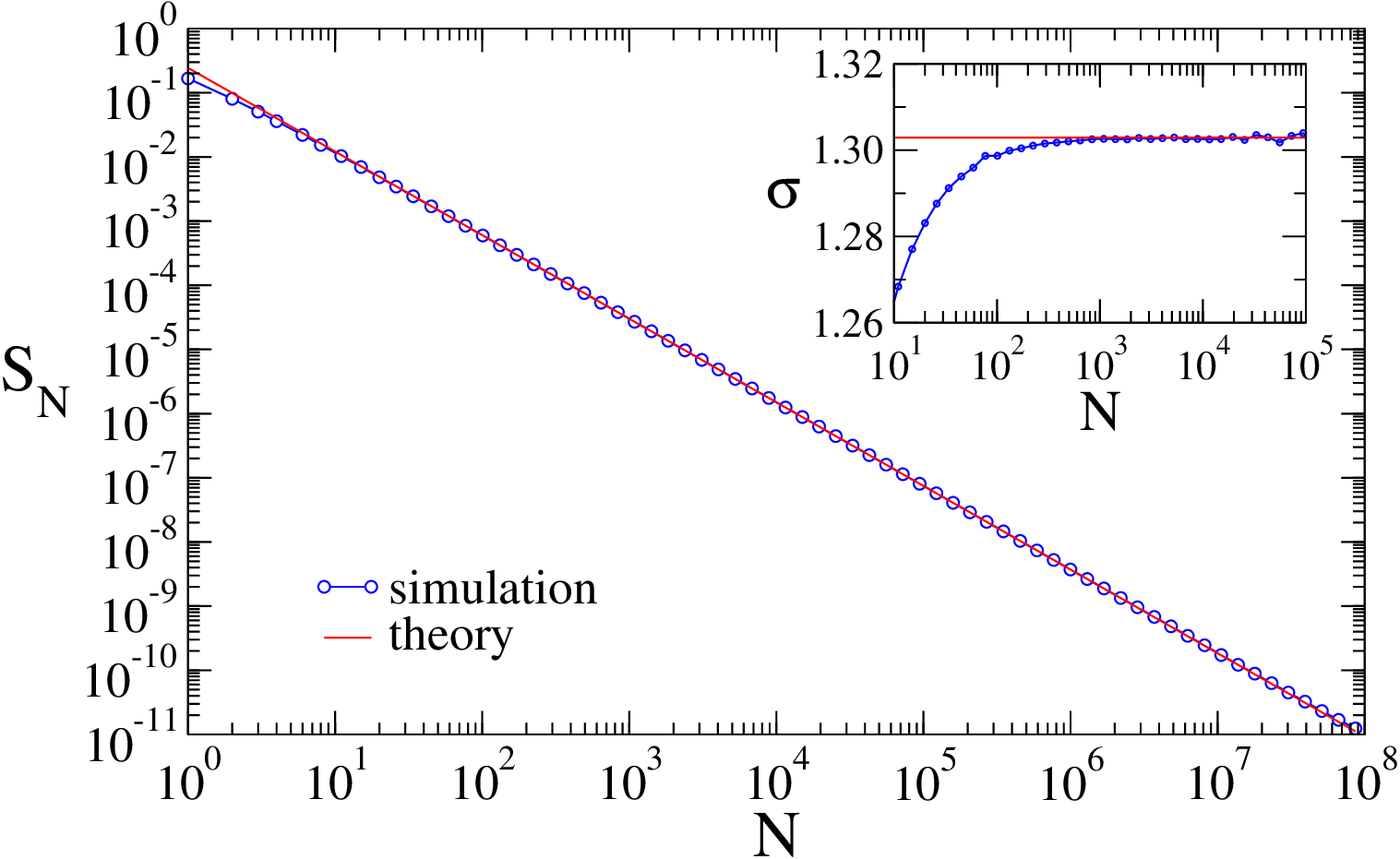}
\caption{The probability $S_N$ versus $N$ for three sequences. Shown
are results of Monte Carlo simulation and the theoretical prediction
\eqref{sigma3}.  The numerical simulation results were obtained from
roughly $10^{13}$ independent runs. The inset shows the local slope
$\sigma\equiv -d\ln S/d\ln N$.}
\label{fig-sn3}
\end{figure}

We now consider three sequences. We denote the third sequence of
random variables as $\{Z_1,Z_2,\ldots,Z_N\}$ and the corresponding
sequence of maxima as $\{z_1,z_2,\ldots,z_N\}$. We are interested in
the probability $S_N$ that the three sequences remain perfectly
ordered as illustrated in Fig.~\ref{fig-order}:
\begin{equation}
\label{condition3}
x_n > y_n > z_n, \qquad  n=1,2,\ldots,N. 
\end{equation}
One immediately finds $S_1=\frac{1}{3!}=\frac{1}{6}$, but it is
challenging to derive $S_N$ for $N\geq 2$ (see Table I).  Our
numerical simulations (see Fig.~\ref{fig-sn3}) show that $S_N$ decays
algebraically,
\begin{equation}
\label{SN-3-decay}
S_N\sim N^{-\sigma},
\end{equation}
for $N\gg 1$ with $\sigma=1.3028\pm 0.0002$.

When there are three sequences, the survival probability $S_N$ no
longer obeys a closed recursion equation such as \eqref{SN-2-rec}. There 
are, however, closed recursion equations for the
probability $P_{N,j}$ that: (i) the maxima are ordered, that is, the
condition \eqref{condition3} holds, and (ii) the number of variables
from the first sequence that are larger than the median maxima equals
$j$ (with $1\leq j\leq N$) as follows 
\begin{equation}
\underbrace{OO\cdots OO}_{3N-1-j}\,Y\underbrace{XX\cdots XX}_{j}.
\end{equation}
In this schematic representation, the $3N$ variables are ordered from
smallest (on the left) to largest (on the right), and moreover, the
labels are ignored. Further, the symbol $O$ stands for a variable from
either one of the three sequences. The number of variables from the
first, second, or third sequence are all equal to $N$. The survival
probability is the sum of the probabilities $P_{N,j}$:
\begin{equation}
\label{SN-sum}
S_N=\sum_{j=1}^N P_{N,j}.
\end{equation}

The probability $P_{N,j}$ obeys the recursion equation 
\begin{eqnarray}
\label{PNj-rec}
P_{N+1,j}
&\!=\!&\frac{3N+2-j}{3N+3}\frac{3N+1-j}{3N+2}\frac{3N-j}{3N+1}\,P_{N,j}\\
&\!+\!&\frac{3N+2-j}{3N+3}\frac{3N+1-j}{3N+2}\frac{j}{3N+1}\,P_{N,j-1}\nonumber\\
&\!+\!&\frac{3N+2-j}{(3N+3)(3N+2)(3N+1)}\sum_{k=j}^{N+1}(3N-k)P_{N,k}\nonumber\\
&\!+\!&\frac{3N+2-j}{(3N+3)(3N+2)(3N+1)}\sum_{k=j}^{N+1}k\,P_{N,k-1}.\nonumber
\end{eqnarray}
To derive \eqref{PNj-rec} we consider the variable $X_{N+1}$ in the
first sequence, the variable $Y_{N+1}$ in the second sequence, and the
variable $Z_{N+1}$ in the third sequence.  The first line in
\eqref{PNj-rec} accounts for the situation where the variable $j$ is
not affected by the introduction of these three variables. (This is
the most likely scenario when $N$ is large.)  The three fractions
account for the probabilities of this event: $X_{N+1}<y_N$ with
probability $\frac{3N-j}{3N+1}$, the probability that $Y_{N+1}<y_N$ is
$\frac{3N+1-j}{3N+2}$, and the probability that $Z_{N+1}<y_N$ is
$\frac{3N+2-j}{3N+3}$.  The next three terms correspond to the three
complementary scenarios, e.g., the second line accounts for the
situation where $X_{N+1}>y_N$ but $Y_{N+1}<y_N$ so that the index $j$
increases, $j\to j+1$.

The recursion \eqref{PNj-rec} is subject to the ``initial'' condition
\hbox{$P_{1,k}=\frac{1}{6}\delta_{k,1}$}. The first iteration of
\eqref{PNj-rec} yields $P_{2,1}=\frac{1}{18}$, $P_{2,2}=\frac{1}{40}$,
and hence, $S_2=\frac{29}{360}$. Table I lists the next few values of
the survival probability, obtained from iteration of the recursion
equation \eqref{PNj-rec}.

\begin{table}[t]
\begin{tabular}{|c|l|l|}
\hline
$N$&$S_N$&$(3N)!\,S_N$\\
\hline
$1$ & $\frac{1}{6}$ & $1$ \\
$2$ & $\frac{29}{360}$ & $58$   \\
$3$ & $\frac{4\, 597}{90\, 720}$ & $18\, 388$  \\
$4$ & $\frac{5\, 393}{149\, 688}$ & $17\, 257\, 600$ \\
$5$ & $\frac{179\, 828\, 183}{6\, 538\, 371\, 840}$ & $35\, 965\, 636\, 600$ \\
$6$ & $\frac{352\, 052\, 449\, 513}{16\, 005\, 934\, 264\, 320}$ & $140\, 820\, 979\, 805\, 200$ \\
\hline
\end{tabular}
\caption{The survival probability $S_N$ for $N=1,2,\ldots,6$. Also
listed are the integers $(3N)!S_N$. These are the number of possible
ways to order the $3N$ random variables $X_n$, $Y_n$, and $Z_n$ such
that the condition \eqref{condition3} holds.}
\end{table}

We are primarily interested in the $N\to\infty$ asymptotic behavior.
In this limit, the probabilities $P_{N,j}$ simplify greatly because
the variable $N$ and $j$ become uncorrelated! Numerical evaluation of
the recursion \eqref{PNj-rec} shows that $P_{N,j}$ factorizes
\begin{equation}
\label{factor}
P_{N,j}\simeq S_N\,p_j.
\end{equation}
The quantity $p_j$ is the limiting rank distribution of the median
record, i.e., the probability that there are $j$ variables exceeding
the median record.  From \eqref{SN-sum} and \eqref{factor} we confirm
that the distribution $p_j$ is properly normalized
\begin{equation}
\label{norm}
\sum_{j=1}^\infty p_j=1.
\end{equation}

\begin{figure}[t]
\includegraphics[width=0.425\textwidth]{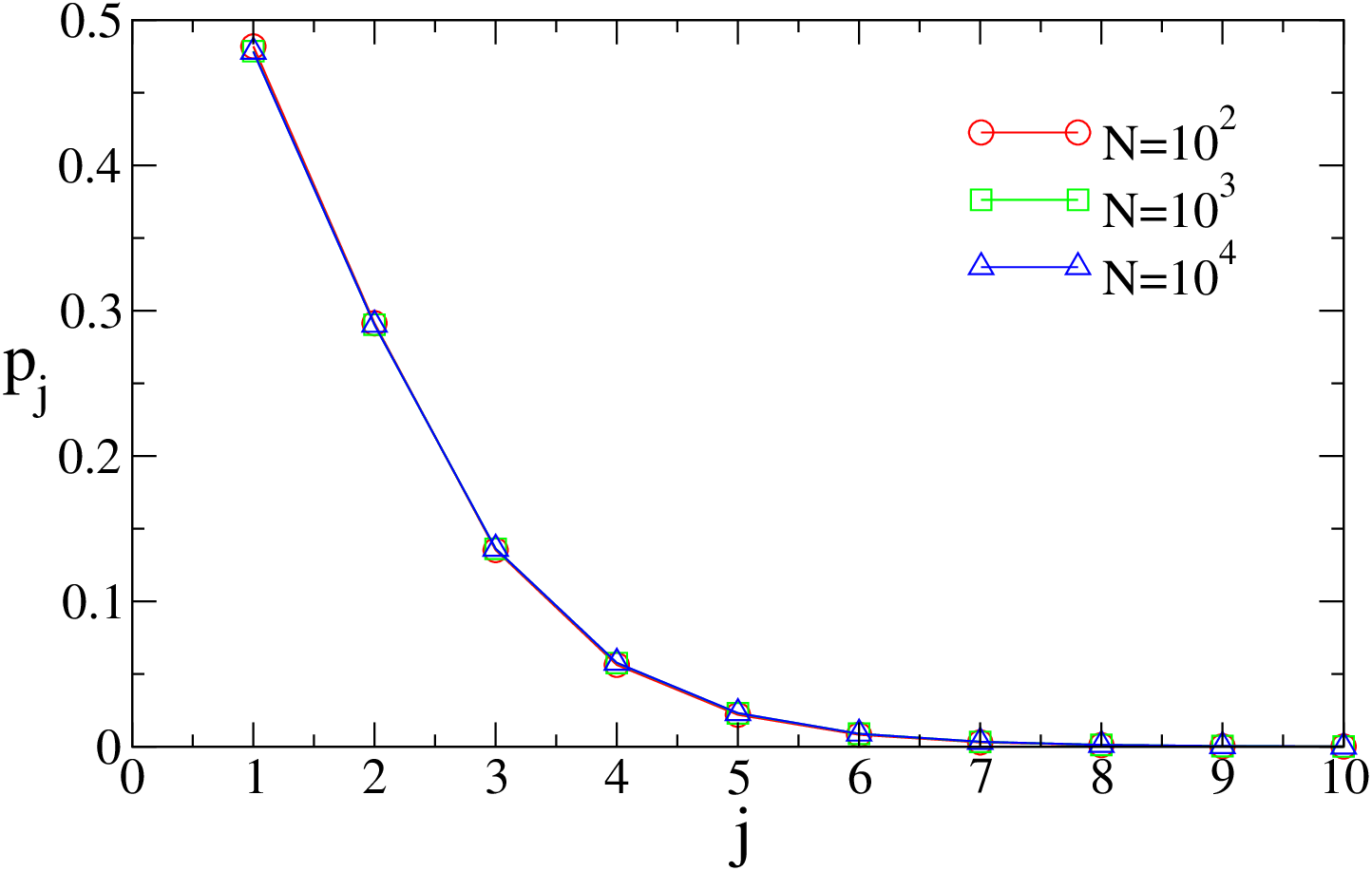}
\caption{The rank distribution $p_j$ versus $j$. Shown are results of
numerical evaluation of the recursion \eqref{PNj-rec} with
increasing values of $N$.}
\label{fig-pj}
\end{figure}

We now substitute \eqref{factor} along with \eqref{SN-3-decay} into
Eq.~\eqref{PNj-rec} and divide both sides by $S_N$. The leading terms
on the right- and left-hand sides [these are of order $O(1)$]
cancel. Evaluating corrections to the leading behavior [these are of
order $O(N^{-1})$], we arrive at the remarkably simple recursion
relation for the rank distribution
\begin{equation}
\label{pj-rec}
\sigma\,p_j=(j+1)\,p_j-\frac{j}{3}p_{j-1}-\frac{1}{3}\sum_{k=j}^\infty p_k.
\end{equation}
In deriving \eqref{pj-rec}, we replaced $S_{N+1}/S_N$ by
$1-\sigma\,N^{-1}$.  The first, second, and third terms on the
right-hand side arise from the first, second, and third terms in
\eqref{PNj-rec}.  The recurrence \eqref{pj-rec}, has to be solved
subject to the normalization \eqref{norm}, and the exponent $\sigma$
is essentially an eigenvalue.

First, we mention a few straightforward results. Summing
\eqref{pj-rec} we express the average rank $\langle j\rangle =\sum_j
j\,p_j$ through the exponent $\sigma$,
\begin{equation}
\label{jav}
\langle j\rangle = 3\sigma-2.
\end{equation}
Next, we iterate \eqref{pj-rec} to obtain the first few probabilities, 
\begin{equation}
\begin{split}
p_1 &= \frac{1}{3(2-\sigma)},\\
p_2 &= \frac{7-3\sigma}{3^2(2-\sigma)(3-\sigma)},\\
p_3 &= \frac{59-48\sigma+9\sigma^2}{3^3(2-\sigma)(3-\sigma)(4-\sigma)}.
\end{split}
\end{equation}

The rank distribution is a rapidly decreasing function
(Fig.~\ref{fig-pj}). To the leading order, the distribution decays
exponentially, $p_j\propto 3^{-j}$, as seen by comparing the leading
terms $j\,p_j=(j/3)p_{j-1}$. The algebraic correction to this leading
asymptotic behavior can be easily extracted from
\eqref{pj-rec}. Indeed, one writes $p_j = 3^{-j}f_j$ and recasts
\eqref{pj-rec} into
$\big(\sigma+\frac{1}{2}\big)f_j=(j+1)f_j-jf_{j-1}$ for $j\gg 1$, from
which we get $f_j\sim j^{\sigma-1/2}$ and hence
\begin{equation}
\label{pj-tail}
p_j\simeq b\, j^{\sigma-1/2}3^{-j}.
\end{equation}
The prefactor $b=1.58063$ is derived in the Appendix (see Appendices A
and B).

To determine the eigenvalue $\sigma$, we employ the generating
function technique. By multiplying \eqref{pj-rec} by $z^{j+1}$ and
summing over all $j\geq 1$, we recast the infinite system of equations
\eqref{pj-rec} into the first-order differential equation
\begin{equation}
\label{P-eq}
(3-z)\frac{dP(z)}{dz}+ P(z)\left(\frac{1}{1-z}-\frac{3\sigma}{z}\right)=\frac{z}{1-z}.
\end{equation}
for the generating function
\begin{equation}
\label{P-def}
P(z) = \sum_{j\geq 1} p_j\, z^{j+1}\,.
\end{equation}

The normalization \eqref{norm} yields $P(1)=1$ and the average rank
\eqref{jav} gives $P'(1)=3\sigma-1$. In addition, we have
$P(0)=0$. The solution to Eq.~\eqref{P-eq} subject to the above
boundary conditions reads
\begin{equation}
\label{Pz-sol}
\begin{split}
P(z) &=
 \sqrt{\frac{1-z}{3-z}}\!\left(\frac{z}{3-z}\right)^\sigma\,U(z),\\
U(z) &= \int_0^z \,\frac{du}{(1-u)^{3/2}}\,\frac{(3-u)^{\sigma-1/2}}{u^{\sigma-1}}\,.
\end{split}
\end{equation}

To obtain the eigenvalue $\sigma$, we evaluate the behavior of this
solution in the vicinity of $z=1$. The solution consists of a regular
term $P_{\text{reg}}(z)$ and a singular term $P_{\text{sing}}(z)$,
that is $P(z)=P_{\text{reg}}(z)+P_{\text{sing}}(z)$.  We already know
that $P(1)=1$ and $P'(1)=3\sigma-1$, and hence, $P_{\rm
reg}(z)=1-(3\sigma-1)(1-z)+\cdots$ as $z\to 1$. On the other hand, the
leading behavior of the singular term is (see Appendix A)
\begin{equation}
\label{Pz-irreg}
P_{\text{sing}}(z)\simeq  F(\sigma)\sqrt{1-z}.
\end{equation}
The amplitude $F(\sigma)$ can be expressed in terms of the Euler gamma
function the hypergeometric function 
\begin{equation}
\label{F}
F(\sigma)\!=\!-\sqrt{\pi}\,\frac{\Gamma(2\!-\!\sigma)}{\Gamma\big(\frac{3}{2}\!-\!\sigma\big)}\,\,
{}_2F_1\!\left(-\tfrac{1}{2}, \tfrac{1}{2}-\sigma; \tfrac{3}{2}-\sigma; -\tfrac{1}{2}\right).
\end{equation}

The quantity $P'(1)$ must be finite and hence, the leading term of the
singular component of the solution must vanish,
$F(\sigma)=0$. Consequently, the exponent $\sigma$ is a root of the
following equation involving hypergeometric function
\begin{equation}
\label{sigma}
{}_2F_1\!\left(-\tfrac{1}{2}, \tfrac{1}{2}-\sigma; \tfrac{3}{2}-\sigma; -\tfrac{1}{2}\right) = 0.
\end{equation}
The quantity $\sigma$ is therefore a transcendental number,
\begin{equation}
\label{sigma3}
\sigma = 1.302931\ldots
\end{equation}
Results of Monte Carlo simulations are in excellent agreement with
this theoretical prediction (Fig.~\ref{fig-sn3}).  In contrast with
the behavior \eqref{SN-2-decay} where the decay exponent is rational,
we see that for three sequences the exponent $\sigma$ governing the
behavior \eqref{SN-3-decay} is apparently irrational.

\subsection{General Case}

We now discuss the general case where there are $m$ sequences, each
containing $N$ random variables. All $m\,N$ random variables are
independently drawn from the same distribution function. Based on the
results for two and three sequences, we expect that the probability
$S_N$ that the $m$ maxima remain ordered is universal, being
independent of the details of the distribution function from which the
variables are drawn. Moreover, we anticipate that the survival
probability decays algebraically,
\begin{equation}
\label{SN-gen-decay}
S_N\sim N^{-\sigma_m}
\end{equation}
for large $N$. Henceforth, the dependence of $S_N$ on $m$ is left
implicit.  The decay exponent $\sigma_m$ depends on the number of
sequences $m$. We already know the values $\sigma_1=0$,
$\sigma_2=1/2$, and $\sigma_3\cong 1.302931$.

\begin{table}[t]
\begin{tabular}{|l|l|}
\hline
$m$&$\sigma_m$\\
\hline
$1$ & $0$                  \\
$2$ & $1/2$         \\
$3$ & $1.302931\ldots$ \\
$4$ & $2.255 \pm 0.015$    \\
$5$ & $3.24 \pm  0.03$     \\
$6$ & $4.2 \pm 0.1 $      \\
$7$ & $5.2 \pm 0.2$        \\
\hline
\end{tabular}
\caption{The exponent $\sigma_m$ versus the number of sequences
  $m$. The numerical simulation results represent an average over
  roughly $10^{13}$ independent realizations.}
\end{table}

\begin{figure}[t]
\includegraphics[width=0.425\textwidth]{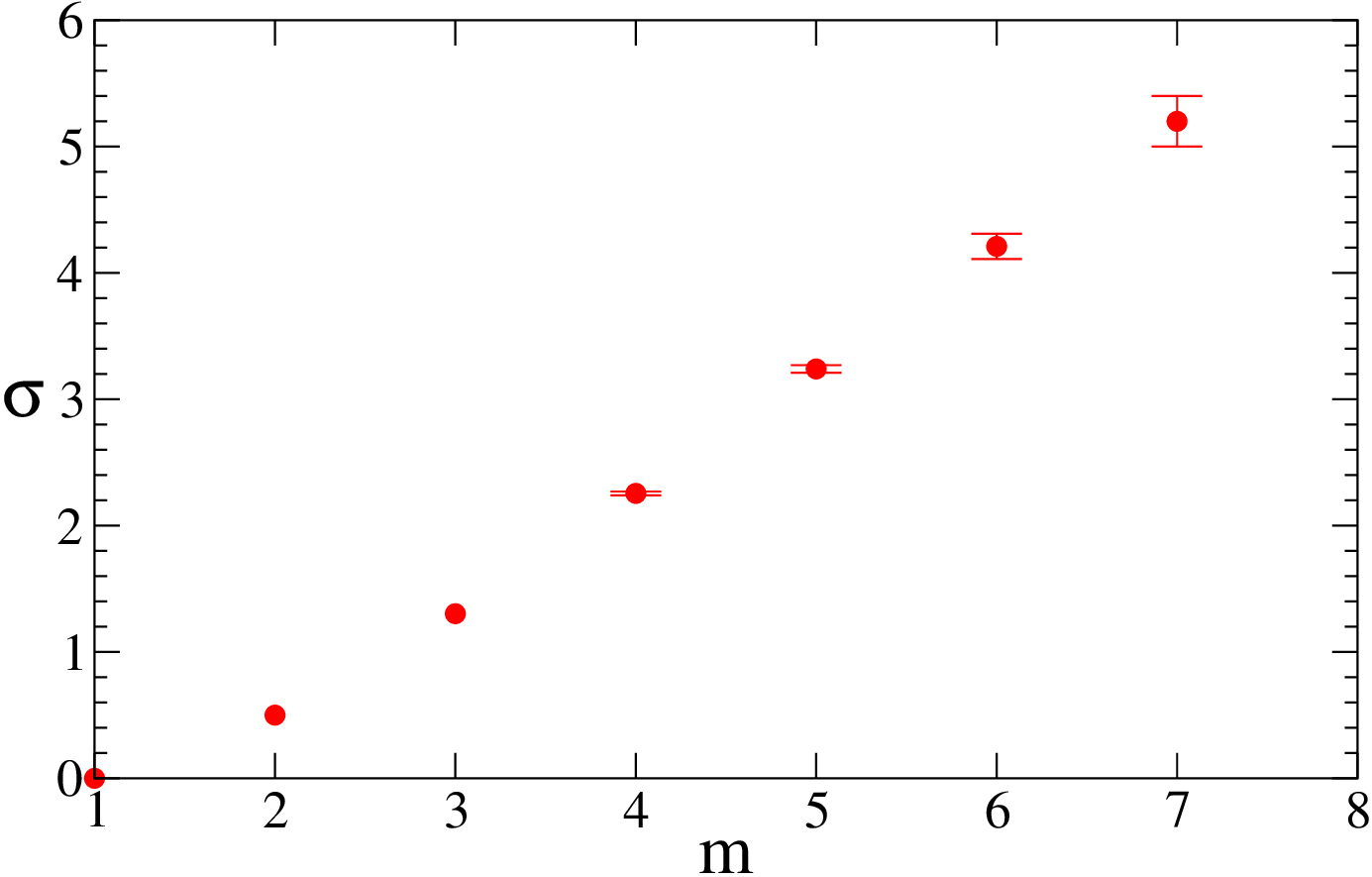}
\caption{The exponent $\sigma$ versus $m$. Shown are results of
numerical simulations, listed in Table I.}
\label{fig-sigmam}
\end{figure}

In principle, the recursive equation \eqref{PNj-rec} for the case
$m=3$ can be generalized to higher values of $m$. However, such a
description involves the positions of all $m-2$ intermediate maxima,
and it is tedious. Instead, we use Monte Carlo simulations.  Table II
lists the numerically obtained values for $4\leq m \leq 7$, while
figure \ref{fig-sigmam} shows the very same data points. These results
suggest that the exponent grows linearly with the number of
sequences. Below, we derive upper and lower bounds for the exponent
and from these two bounds, we deduce the linear growth analytically.

It is straightforward to establish a lower bound for the probability
$S_N$ and consequently, an upper bound for the exponent
$\sigma_m$. Let us consider the special case where: (i) the $m$ maxima
are properly ordered on the very first step, and (ii) all $m$ maxima
remain constant. The probability for the first event is $1/m!$ and the
probability for the second scenario is $N^{-m}$. Hence, we have the
lower bound
\begin{equation}
\label{SN-lower}
S_N\geq \frac{1}{m!}\,N^{-m}\, .
\end{equation}
This simple argument gives the upper bound $\sigma_m \leq m$.

To establish an upper bound for the probability $S_N$ and a consequent
lower bound for the exponent $\sigma_m$, we introduce a natural
generalization of the survival probability $S_N$ for the case $m=2$.
Equation \eqref{SN-2-sol} corresponds to the situation where there are
two sequences of random variables and the first set of maxima is
always larger. When there are $m$ sequences, we can similarly ask:
what is the probability $A_N$ that the first set of maxima is always
ahead? For example, when $m=3$ this is the probability that $x_n>y_n$
and $x_n>z_n$ for all $n=1,2,\ldots,N$. The probability $A_N$
satisfies a straightforward generalization of the recursion
\eqref{SN-2-rec}
\begin{equation}
\label{AN-rec}
A_{N}=A_{N-1}\left(1-\frac{m-1}{m\,N}\right).
\end{equation}
When $m$ new variables are added, the probability that a new
``global'' maximum is set equals $1/N$ and the probability that this
new maximum belongs to one of the $m-1$ trailing sequences is simply
$\frac{m-1}{m}$. Starting from $A_0=1$, the recursion equation
\eqref{AN-rec} gives
\begin{equation}
\label{AN-sol}
A_{N}=\frac{\Gamma(N+\tfrac{1}{m})}{\Gamma(\tfrac{1}{m})\,\Gamma(N+1)},
\end{equation}
where $\Gamma(x)$ is the Euler Gamma function.  When \hbox{$m=2$}, we
recover $S_N$ given in \eqref{SN-2-sol}. Using the asymptotic
relation, $\Gamma(N+a)/\Gamma(N)\simeq N^a$ as $N\to\infty$, we obtain
the power-law decay
\begin{equation}
\label{AN-decay}
A_{N}\simeq \frac{1}{\Gamma(\tfrac{1}{m})}\, N^{-\alpha_m}, \qquad 
\alpha_m=1-\frac{1}{m},
\end{equation}
for large $N$. The family of exponents $\alpha_m$ approaches a
constant $\alpha\to 1$ in the limit $m\to\infty$,

\begin{figure}[t]
\includegraphics[width=0.425\textwidth]{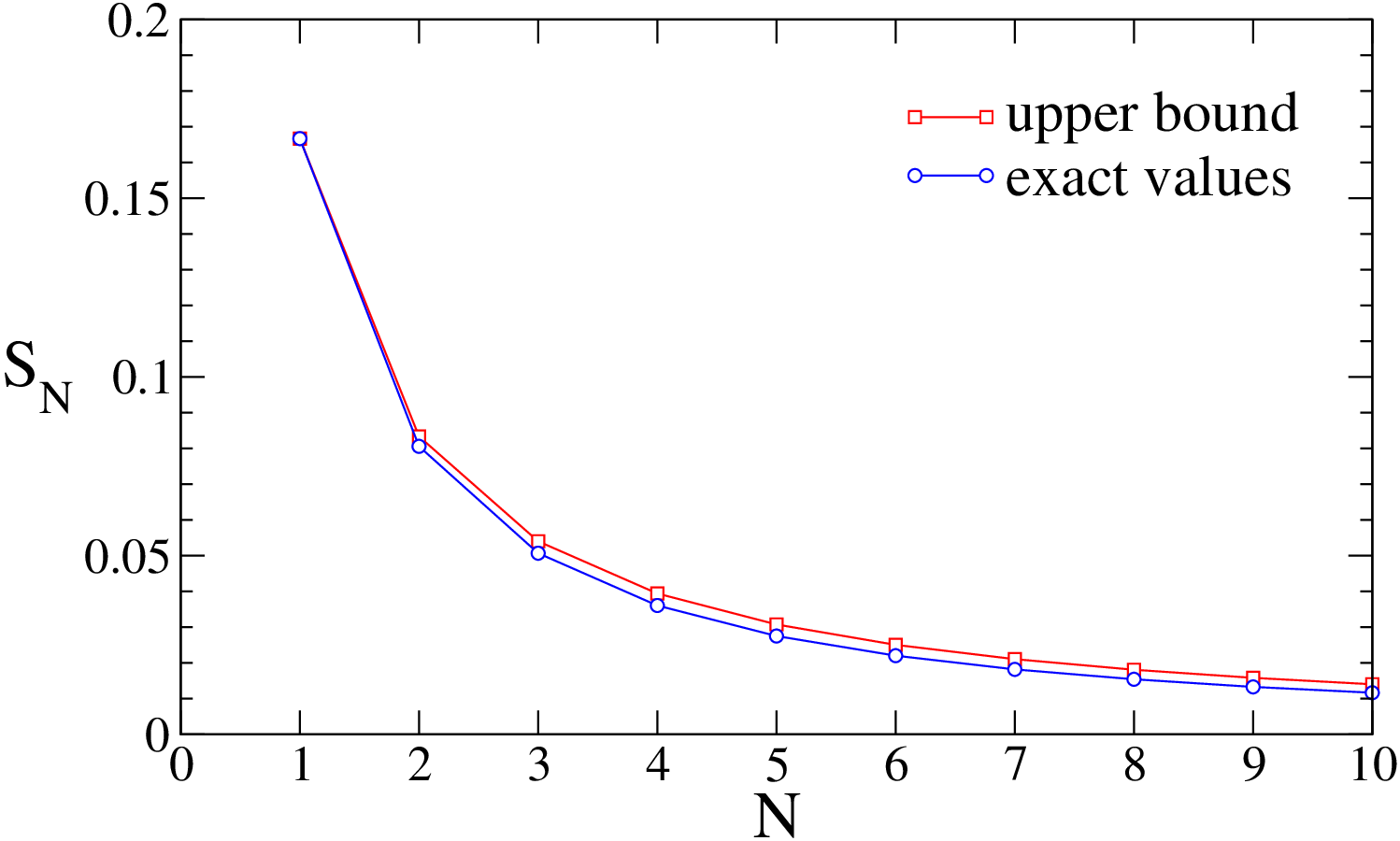}
\caption{The survival probability $S_N$ for $m=3$.  The upper bound
  \eqref{SN-upper} shown in squares is compared with the exact values
  (circles), obtained from the exact recursion \eqref{PNj-rec}.}
\label{fig-sn}
\end{figure}

We are now in a position to construct a lower bound for $S_N$. The
probability that the first maximum is always larger than all other
$m-1$ maxima is given by $A_N$ in \eqref{AN-sol}. The probability that
the second maximum is always larger then $m-2$ remaining maxima can be
estimated by replacing $m$ with $m-1$ in \eqref{AN-sol}. Similarly,
the probability that the third maximum is always larger than the next
$m-3$ maxima is approximated by replacing $m$ with $m-2$ in
\eqref{AN-sol}, and so on. The product of these $m-1$ probabilities
constitutes an upper bound
\begin{equation}
\label{SN-upper}
S_N\leq \prod_{k=2}^m 
\frac{\Gamma(N+\tfrac{1}{k})}{\Gamma(\tfrac{1}{k})\,\Gamma(N+1)}.
\end{equation}
Figure \ref{fig-sn} demonstrates that this upper bound gives a very
good approximation for the probability $S_N$. For example, when $m=3$,
the upper bound is $\frac{1}{12}=\frac{30}{360}$ whereas the exact
value is $S_2=\frac{29}{360}$ (see Table I). The lower bound
\eqref{SN-lower} is a much poorer approximation, in comparison.

By using the upper bound \eqref{SN-upper} and the asymptotic behavior
\eqref{AN-decay} we arrive at the lower bound,
\begin{equation}
\label{sigma-lowe}
\sigma_m\geq \alpha_2+\alpha_3+\cdots\alpha_m\,.
\end{equation}
Hence, we conclude that the exponent $\sigma_m$ is bounded from above
and from below as follows, 
\begin{equation}
\label{sigma-bounds}
m-\left(1+\frac{1}{2}+\frac{1}{3}+\cdots+\frac{1}{m}\right) \leq \sigma_m \leq m \,.
\end{equation}
For example, for $m=3$ we have the bounds $\tfrac{7}{6}\leq \sigma_3
\leq 3$, and indeed, the lower bound is much tighter compared with the
upper bound. Most significantly, the two bounds establish the linear
growth (see figure \ref{fig-sigmam})
\begin{equation}
\label{sigma-linear}
\sigma_m\simeq m.
\end{equation}
The asymptotic behavior $1+\frac{1}{2}+\cdots+\frac{1}{m}\simeq \ln
m+\gamma$ (here $\gamma\cong 0.577215$ is the Euler constant) shows
that the deviation from linear growth is at most logarithmic. The
linear growth \eqref{sigma-linear} is in contrast with the quadratic
$m(m-1)/4$ growth of the exponent characterizing ordering of random
walks \cite{mef,hf,G99,bjmkr,bk-mult}. Finally, we mention a numerical
observation: the empirical formula $\tilde\sigma_m=(m-1)^2/m$, which
is exact for $m=1$ and $m=2$, yields an excellent approximation for
the values listed in Table I.

The exponents $\sigma_m$ provide an excellent approximation to an
analogous set of exponents that characterize random walks.  Let $s_N$
be the probability that the maxima of the positions of $m$ independent
random walks, each consisting of $N$ steps, are ordered. This quantity
decays algebraically $s_N\sim N^{-\nu_m}$.  For two random walks, it
was established analytically that $\nu_2=\frac{1}{4}$ \cite{bk}. In
view of the identity $\nu_2=\sigma_2/2$, we compare the values
$\sigma_m/2$ with the values of $\nu_m$ obtained using numerical
simulations for $m=3,4,5,6$ \cite{bk}. Table III shows that the
ordering exponents, obtained in the present study for uncorrelated
random variables, provide an excellent approximation for the ordering
exponents that characterize the maximal positions of random walks, at
least for small values of $m$. Random walk positions are certainly
correlated random variables, and therefore, the results above, which
are strictly valid for uncorrelated random variables, may in practice
provide useful approximations for certain classes of correlated random
variables.

We also mention that the quantity $s_N$ is not universal as it
depends, albeit rather weakly, on the step length distribution
\cite{bk-unpub}.  Nevertheless, the decay exponents $\nu_m$ listed in
Table III are universal, namely they are valid as long as the step
length distribution is symmetric and has finite variance \cite{bk}.

\begin{table}[t]
\begin{tabular}{|l|l|l|}
\hline
$m$ &  $\nu_m$ & $\sigma_m/2$  \\
\hline
$2$ & $1/4$    & $1/4$         \\
$3$ & $0.653$  & $0.651465$    \\
$4$ & $1.13$   & $1.128$       \\
$5$ & $1.60$   & $1.62$        \\
$6$ & $2.01$   & $2.10$         \\
\hline
\end{tabular}
\caption{The exponent $\nu_m$ for ordering of random walk maxima
\cite{bk} versus the values $\sigma_m/2$, see Table II.}
\end{table}

\section{Partial Order}

The exponents $\sigma_m$ characterizing the statistics of perfectly
ordered maxima are part of a broader family of exponents. We have
already obtained one such family of exponents,
$\alpha_m=1-\frac{1}{m}$, that characterize the probability $A_N$ that
the first sequence of maxima is always larger than all other $m-1$
maxima. For example, when $m=2$ the requirement that $x_n$ is largest
is equivalent to the requirement $x_n>y_n$ for all
$n=1,2,\ldots,N$. Therefore, $S_N=A_N$ for $m=2$, and hence
$\alpha_2=\sigma_2$. Another trivial identity is $\alpha_1=\sigma_1$.

Similarly, we can introduce the probability $B_N$ that the first two
sets of maxima remain ordered and exceed the other $m-2$ maxima. In
other words, the sequence of maxima $\{x_1,x_2,\ldots,x_N\}$ is always
the largest, and the sequence $\{y_1,y_2,\ldots,y_N\}$ is always the
second largest. Therefore, the probabilities $B_N$ and $S_N$ are
identical when there are two or three sequences. We expect the
algebraic decay
\begin{equation}
\label{BN-decay}
B_N\sim N^{-\beta_m},
\end{equation}
with $\beta_2=\sigma_2$ and $\beta_3=\sigma_3$.

\begin{figure}[t]
\includegraphics[width=0.425\textwidth]{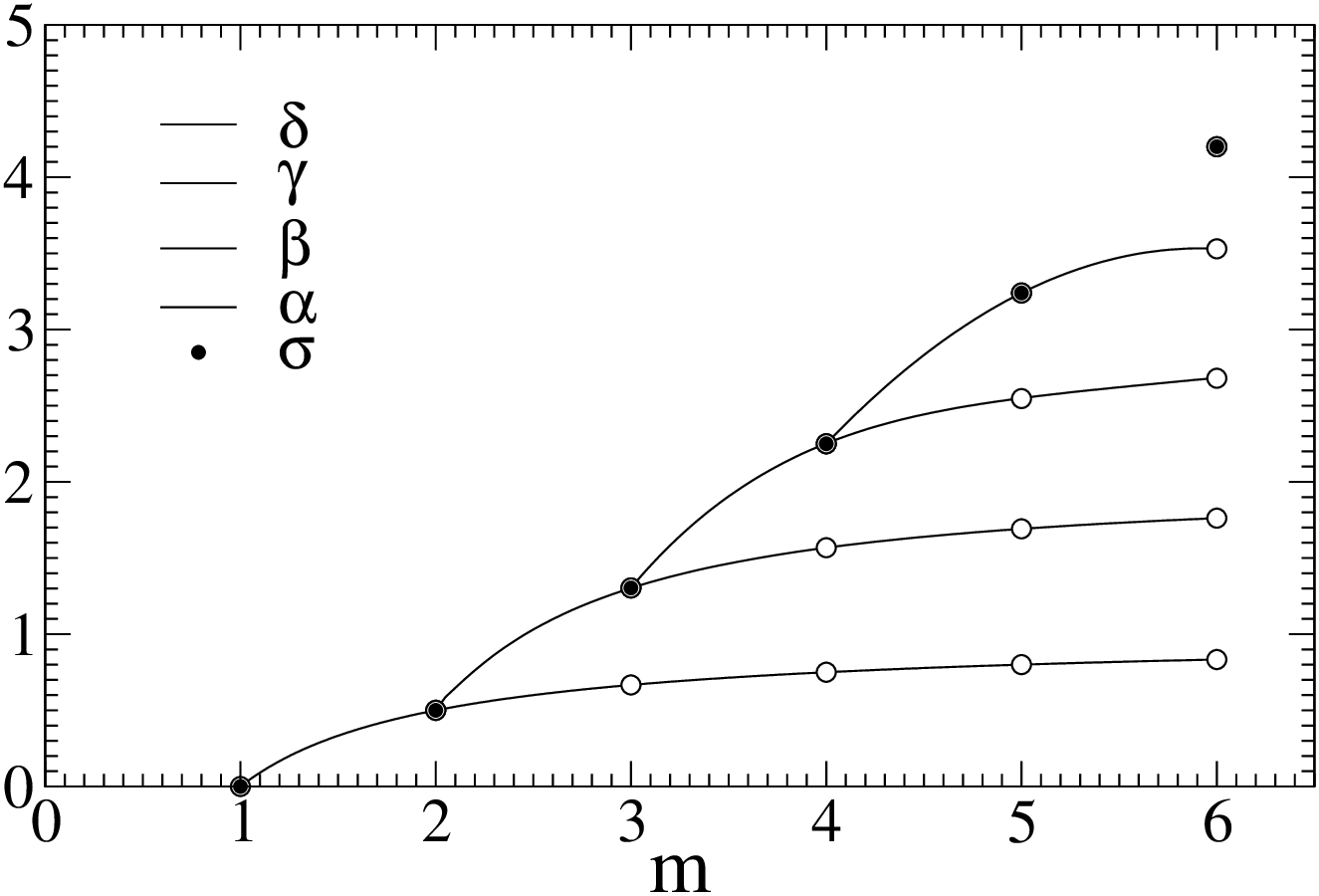}
\caption{Families of ordering exponents. Shown are the first four
  families of exponents; the values of $\alpha_m$, $\beta_m$,
  $\gamma_m$, and $\delta_m$ are listed in Table III (circles).  The
  ordering exponents $\sigma_m$ describing perfectly ordered maxima
  (bullets) are also shown. The analytic curves for $\alpha$ and
  $\beta$ are given in equations \eqref{AN-decay} and
  \eqref{beta}. The curves for $\gamma$ and $\delta$ represent a
  fourth-order polynomial best fit to the simulation results.}
\label{fig-family}
\end{figure}

The recursion equation \eqref{PNj-rec} is straightforward to
generalize from three to $m\geq 2$ sequences, thereby giving a
recursive calculation of the probability $B_N$. By repeating the
analysis leading to \eqref{PNj-rec} we find that the rank distribution
$p_j$ obeys the recursion equation
\begin{equation}
\label{pj-rec-1}
\beta\,p_j=(j+1)\,p_j-\frac{j}{m}p_{j-1}-\frac{1}{m}\sum_{k=j}^\infty p_k
\end{equation}
and the normalization condition \eqref{norm}.  The exponent
$\beta\equiv \beta_m$ is again an eigenvalue, and the average rank can
be expressed through this exponent, \hbox{$\langle j\rangle =
[1+m(\beta-1)]/(m-2)$}.  Moreover, the rank distribution has the
following tail \hbox{$p_j\sim j^{\beta-(m-2)/(m-1)}\,m^{-j}$}. By
following the analysis in Sec.~\ref{sec:3}, it is straightforward to
show that exponent $\beta$ is a root of a transcendental equation
involving the hypergeometric function,
\begin{equation}
\label{beta}
{}_2F_1\!\left(-\mu, 1-\mu-\beta; 2-\mu-\beta; -\mu \right) = 0,
\end{equation}
with the shorthand notation $\mu=1/(m-1)$.  When $m=3$ this equation
coincides with \eqref{sigma} and hence $\beta_3=\sigma_3$.  The next
three values are $\beta_4=1.56479$, $\beta_5=1.69144$, and
$\beta_6=1.76164$.  One can also deduce the asymptotic behavior,
$2-\beta_m\simeq m^{-1}$, when $m\gg 1$. Just like the family of
exponents $\alpha_m$, the curve $\beta_m$ saturates in the large $m$
limit: $\beta_m\to 2$. While the parameter $m$ is discrete, the
solution to the transcendental equation \eqref{beta} can be evaluated
for all $m\geq 2$ and the resulting {\it continuous} curve is shown in
Fig.~\ref{fig-family}.

There is an infinite set of probabilities generalizing $A_N$ and
$B_N$: The probability $C_N$ that the first three maxima (out of a
total of $m\geq 3$) remain ordered $x_n>y_n>z_n$ and exceed all others
for all $1\leq n\leq N$; the probability $D_N$ that the first four
maxima (out of total of $m\geq 4$) remain ordered and exceed all
others, and so on.  These probabilities are characterized by three
families of exponents,
\begin{equation}
\label{CD-decay}
C_N\sim N^{-\gamma_m},
\quad
D_N\sim N^{-\delta_m}.
\end{equation}
We have $\beta_3=\gamma_3=\sigma_3$, $\gamma_4=\delta_4=\sigma_4$, and
$\delta_5=\sigma_5$. The first four families of exponents are shown in
Fig.~\ref{fig-family}. These families of exponents form an intriguing
structure that resembles a scallop. The smallest two integer exponents
in the $m$th family coincide with the exponents $\sigma_m$ and
$\sigma_{m+1}$.  An interesting question for future research is
whether the families of ordering exponents adhere to a universal
scaling curve when the number of conditions such as \eqref{condition3}
becomes very large.

\begin{table}[t]
\begin{tabular}{|l|l|l|l|l|}
\hline
$m$&$\alpha$&$\beta$&$\gamma$&$\delta$\\
\hline
$1$ & $0$    &                      & &             \\
$2$ & $1/2$ &  $1/2$           & &             \\
$3$ & $2/3$ &  $1.302931$ & $1.302931$   &    \\
$4$ & $3/4$ &  $1.56479$   & $2.255$         &  $2.255$   \\
$5$ & $4/5$ &  $1.69144$   & $2.547$         &  $3.24$      \\
$6$ & $5/6$ &  $1.76164$   & $2.680$         &  $3.53$      \\
\hline
\end{tabular}
\caption{The first few families of exponents. The exponents $\alpha_m$
and $\beta_m$ are obtained analytically. When $m\geq 4$, the exponents
$\gamma_m$ and $\delta_m$ are from numerical simulations.}
\end{table}

\section{Conclusions}

We have shown that the probability that the running maxima of
independent sets of random variables are ordered decays algebraically
with the number of variables. The scaling exponents that characterize
this decay are in general nontrivial. When there are three sequences,
the scaling exponent is eigenvalue of a recurrence equation, and it is
also a root of a transcendental equation. The scaling exponents grow
linearly with the number of independent sequences. We have also seen
that ordering exponents for uncorrelated random variables provide an
excellent approximation for the corresponding set of exponents that
characterize maximal positions of random walks.

The key observation that allowed us to treat the three-variable case
analytically is that the rank of the median maxima decouples from the
sequence length in the asymptotic regime.  This observation, combined
with the power-law decay of the overall survival probability reduces
the complexity of the underlying combinatorial problem: enumerating
the number of ways to order the random variables such that the
respective maxima remain ordered.

One can study the probability that the running maxima of the first $k$
sequences are ordered and exceed the maxima of remaining $m-k$
sequences. We have examined such probabilities for $k=1,2,3,4$ and
derived analytic expressions for the exponents for the case with
persistent leader ($k=1$) and the case with persistent leader and
persistent second leader.  We have seen that there are families of
exponents, in some cases equivalent to eigenvalues, that form an
intriguing structure. For uncorrelated random variables, the
intersection points of these eigenvalue families mark a linear
envelope. It would therefore be interesting to investigate the
corresponding exponent families for ordered maxima of Brownian
trajectories, where that envelope shows quadratic growth
\cite{mef,hf,G99,bjmkr,bk-mult}.

Furthermore, there are many similar survival probabilities, for example, the
probability $L_N$ that the running maxima of one sequence are never
the smallest.  Numerically, we observe the algebraic decay $L_N\sim
N^{-\lambda_m}$.  The first exponent is obvious, $\lambda_2=1/2$. Our
numerical simulations yield a slowly decreasing set of exponents:
$\lambda_3=0.3801$, $\lambda_4=0.3145$, $\lambda_5=0.2726$, and
$\lambda_6=0.2430$.

We emphasize that the probability that the actual random variables
remain perfectly ordered decays exponentially while the probability
that the running maxima maintain perfect order decays much more
slowly, namely algebraically, with sequence length. Hence, it is far
more likely to observe ordered maxima. In several contexts such as
temperature records \cite{whk} or stock market time series \cite{bp},
record highs or record lows are followed very closely to see for
example if one year is the hottest or if one stock is consistently
outperforming its peers. Hence, we expect that the questions we
investigated theoretically in this study may be of practical relevance
in analysis of time series. Moreover, consistently ordered extreme
values provide a natural way to quantify persistent upward or downward
trends in the data.

We anticipate that the actual survival probabilities, particularly the
values of the scaling exponents, may be measurable in empirical data,
including those with correlated random variables. One such example is
inter-event times for earthquakes where a series of recent studies
demonstrate how ``persistence'' properties of maxima of uncorrelated
variables provide excellent predictions for empirical observations
\cite{win,bk13,mb}. 

\bigskip
We acknowledge financial support through US-DOE grant
DE-AC52-06NA25396 for support (EB \& NWL).

\appendix
\onecolumngrid
\section{The generating function $P(z)$ in the limit $z\to 1$}

We evaluate the asymptotic behavior of the function $U(z)$ which appears 
in \eqref{Pz-sol} as $z\to 1$ using the following steps:
\begin{eqnarray}
U(z) &=& \int_0^z \,\frac{du}{(1-u)^{3/2}}\,\frac{(3-u)^{\sigma-1/2}}{u^{\sigma-1} } \\
&=&
\int_0^z \,\frac{du}{(1-u)^{3/2}}\left[\frac{(3-u)^{\sigma-1/2}}{u^{\sigma-1}}-2^{\sigma-1/2}\right]
-2^{\sigma+1/2}+ 2^{\sigma+1/2}(1-z)^{-1/2}\nonumber\\
&=& 
2^{\sigma+1/2}\left(\frac{1}{\sqrt{1-z}} -1\right) + \int_0^1 \,\frac{du}{(1-u)^{3/2}}\left[\frac{(3-u)^{\sigma-1/2}}{u^{\sigma-1}}-2^{\sigma-1/2}\right]
-\int_z^1 \,\frac{du}{(1-u)^{3/2}}\left[\frac{(3-u)^{\sigma-1/2}}{u^{\sigma-1}}-2^{\sigma-1/2}\right]\nonumber\\
&=& 
2^{\sigma+1/2}\left\{\frac{1}{\sqrt{1-z}} -1+ \frac{1}{2}\int_0^1 \,\frac{du}{(1-u)^{3/2}}\left[u^{1-\sigma}\left(\frac{3-u}{2}\right)^{\sigma-1/2}-1\right]-\frac{6\sigma-5}{4}\,\sqrt{1-z}+O\left[(1-z)^{3/2}\right]\right\}\nonumber\\
&=& 2^{\sigma+1/2}\left( \frac{1}{\sqrt{1-z}}+ F(\sigma) - \frac{6\sigma-5}{4}\, \sqrt{1-z} \right)
+O\left[(1-z)^{3/2}\right].\nonumber
\end{eqnarray}
The quantity $F(\sigma)$ can be expressed in terms of the hypergeometric function, 
\begin{eqnarray}
\label{Fsigma}
F(\sigma)
&=&
\frac{1}{2}\int_0^1 \frac{du}{(1-u)^{3/2}}\!\left[u^{1-\sigma}\left(\frac{3-u}{2}\right)^{\sigma-1/2}\!-1\right]-1\\
&=&\frac{1}{2}\int_0^1 dv\,v^{-3/2}\!\left[(1-v)^{1-\sigma}(1+\tfrac{1}{2}v)^{\sigma-1/2}-1\right]-1\nonumber\\
&=&\frac{1}{2}\frac{\Gamma(-\tfrac{1}{2})\,\Gamma(2-\sigma)}{\Gamma(\tfrac{3}{2}-\sigma)}
{}_2F_1(-\tfrac{1}{2},\tfrac{1}{2}-\sigma;\tfrac{3}{2}-\sigma;-\tfrac{1}{2})
-\frac{1}{2}\int_0^1 dv\,v^{-3/2}-1\nonumber\\
&=&-\sqrt{\pi}\,\frac{\Gamma(2-\sigma)}{\Gamma\big(\frac{3}{2}-\sigma\big)}\,\,
{}_2F_1(-\tfrac{1}{2},\tfrac{1}{2}-\sigma;\tfrac{3}{2}-\sigma;-\tfrac{1}{2}).\nonumber
\end{eqnarray}
First, we note that the integral which specifies $F(\sigma)$ is
finite.  In deriving the third line we used the integral representation of
the hypergeometric function
\begin{equation}
\int_0^1 dv\, v^{b-1}(1-v)^{c-b-1}(1-zv)^{-a} = \frac{\Gamma(b)\,\Gamma(c-b)}{\Gamma(c)}\,{}_2F_1(a,b;c;z),
\end{equation}
and the identity ${}_2F_1(a,b;c;z)={}_2F_1(b,a;c;z)$.  In deriving the
fourth line, we used the identity (valid in the sense of
distributions, or generalized functions \cite{gs}) $\int_0^1 dv\,
v^\lambda = \frac{1}{\lambda+1}$. This identity is certainly valid in
the complex plane where $\text{Re}(\lambda)> -1$, and then it may be
analytically continued into the whole complex plane, in particular to
the value $\lambda=-3/2$.

\section{The generating function $P(z)$ in the limit $z\to 3$}

The integral representation \eqref{Pz-sol} implicitly assumes that
$z<1$. To obtain the asymptotic behavior of the generating function
$P(z)$ in the $z\to 3$, we write the solution of Eq.~\eqref{P-eq} in a
different form 
\begin{equation}
\label{PUz}
\begin{split}
P(z) &=  \sqrt{\frac{z-1}{3-z}}\!\left(\frac{z}{3-z}\right)^\sigma\,\tilde U(z)\, , \\
\tilde U(z) &=  \tilde U(3) + \int_z^3 \,\frac{du}{(u-1)^{3/2}}\,\frac{(3-u)^{\sigma-1/2}}{u^{\sigma-1}}\, .
\end{split}
\end{equation}
To determine the constant $\tilde U(3)$ we transform $\tilde U(z)$ as follows:
\begin{eqnarray}
%\label{Uz-long}
\tilde U(z) &=&  \tilde U(3) + \int_z^3 \,\frac{du}{(u-1)^{3/2}}\left[\frac{(3-u)^{\sigma-1/2}}{u^{\sigma-1}}-2^{\sigma-1/2}\right]+
2^{\sigma-1/2}\int_z^3 \,\frac{du}{(u-1)^{3/2}} \nonumber\\
&=&\tilde U(3)  + \int_z^3 \,\frac{du}{(u-1)^{3/2}}\left[\frac{(3-u)^{\sigma-1/2}}{u^{\sigma-1}}-2^{\sigma-1/2}\right]+
2^{\sigma+1/2}\left[\frac{1}{\sqrt{z-1}}-\frac{1}{\sqrt{2}}\right] \nonumber \\
&=& 
\frac{2^{\sigma+1/2}}{\sqrt{z-1}}
+\tilde U(3)-2^\sigma
+\int_1^3 \,\frac{du}{(u-1)^{3/2}}\left[\frac{(3-u)^{\sigma-1/2}}{u^{\sigma-1}}-2^{\sigma-1/2}\right]
-\int_1^z \,\frac{du}{(u-1)^{3/2}}\left[\frac{(3-u)^{\sigma-1/2}}{u^{\sigma-1}}-2^{\sigma-1/2}\right]\nonumber\\
&=&\frac{2^{\sigma+1/2}}{\sqrt{z-1}}
-\int_1^z \,\frac{du}{(u-1)^{3/2}}\left[\frac{(3-u)^{\sigma-1/2}}{u^{\sigma-1}}-2^{\sigma-1/2}\right].\nonumber
\end{eqnarray}
The leading singular contribution to the $P(z)$ equals the sum of the
second, third, and fourth terms on the third line, and these terms
cancel so that the generating function is regular at $z=1$. In the
limit $z\to 3$ we have $P(z)\simeq \sqrt{2}\, 3^\sigma\, \tilde U(3)\,
(3-z)^{-\sigma-1/2}$. In addition, we use the expansion
$(1-x)^{-a}=\sum_{n\geq 0} \Gamma(a+n)\,x^n /[\Gamma(a)\Gamma(n+1)]$
to deduce the tail behavior \eqref{pj-tail} with
\begin{equation}
\label{b}
b = \sqrt{\frac{2}{3}}\,\,\frac{1}{3\,\Gamma(\sigma+1/2)} \left\{2^\sigma
- \int_1^3 \,\frac{du}{(u-1)^{3/2}}\left[\frac{(3-u)^{\sigma-1/2}}{u^{\sigma-1}}-2^{\sigma-1/2}\right] \right\}.
\end{equation}

\end{document}